\documentclass[a4paper,12pt]{article}
\usepackage[utf8]{inputenc}
\usepackage{cancel}
\usepackage{ulem}
\usepackage{amsfonts}
\usepackage{amssymb}
\usepackage{graphicx}
\usepackage{amsmath}
\usepackage{enumerate}
\usepackage{here}
\usepackage{subfloat}
\usepackage{subfig}
\usepackage{color}
\usepackage{float}
\usepackage{here}
\usepackage{cite}
\usepackage{mathrsfs}
\usepackage{float,epsfig}
\usepackage{dcolumn}
\usepackage{multirow}
\usepackage{placeins}
\usepackage{graphicx}
\usepackage{bm}
\usepackage{amsmath,amssymb,amsthm}
\usepackage[colorlinks=true,linkcolor=blue]{hyperref}
\hypersetup{citecolor=magenta}
\usepackage[table]{xcolor}
\title{\bf \Large  
Probing correlation between photon orbits and phase structure  of charged AdS black hole in massive gravity background}

\author{ M. Chabab$^{1}$\footnote{mchabab@uca.ac.ma (Corresponding author) }, H. El Moumni$^{1,2}$\footnote{hasan.elmoumni@edu.uca.ma}, S. Iraoui$^{1}$\footnote{samir.iraoui@ced.uca.ma}, K. Masmar$^{1}$\footnote{karima.masmar@edu.uca.ac.ma}\\
	\\ 
	{\small $^{1}$ High Energy and Astrophysics Laboratory, Physics Department, FSSM, Cadi Ayyad University,
	}\\
	{\small  P.O.B. 2390 Marrakech, Morocco.
	}\\
	{\small $^{2}$ EPTHE, Physics Department, Faculty of Science,  Ibn Zohr University, Agadir, Morocco. }
}

\date{}
\begin{document} 
\maketitle
	
\begin{abstract}

The phase structure of charged Anti-de Sitter black hole in massive gravity is investigated using the unstable circular photon orbits formalism, concretely we establish a direct link between the null geodesics and the critical behavior thermodynamic of such black hole solution.  Our analysis reveals that the radius and the impact parameter corresponding to the unstable circular orbits  can be used to probe the  thermodynamic phase structure. We also show that the latter are key quantities to characterize the order of van der Waals-like phase transition. Namely, we found a critical exponent around $ \delta = 1/2 $.
All these results support further that the photon trajectories can be used as a useful and crucial tool to probe the
thermodynamic black holes criticality.

\end{abstract}

          \tableofcontents
  
\section{Introduction}

Our investigation is loosely motivated by the statement that the information encoded in a black hole can well be uncovered  by observing the matter behavior in the vicinity  of its outsider horizon. Thus,  the black hole identification is directly linked to the analysis of the particle motions that are affected by the strong gravity field near such astrophysical object. There exists at present a quite  big number of studies on geodesics around black holes in various theories of gravity (see \cite{Chandrasekhar,shapiro,Bardeen,Cardoso,Pugliese} and reference therein). Likewise, many years ago, it was shown  that geodesics of a test particle can play a crucial role in our comprehension of some observational effects such as the gravitational deflection angle of light and time delays \cite{Rindler,Sereno,Arakida}. \\

Recently, there has been a significant increasing interest on alternative gravity models  to Einstein theory. One of the popular models is the massive gravity. This theory can be considered as massive spin-2 graviton theory. Historically, the construction of linear theory of massive gravity goes back to Fierz and Pauli, who have developed the ghost free theory of noninteracting massive graviton and added the interaction terms at linearized level of the general relativity. They observed that the mass term must be of the form $m_{g}^{2}\left(h^{2}-h_{\mu\nu}h^{\mu\nu}\right)$ \cite{Fierz1}. The theory of Fierz and Pauli suffers from the problem of van Dam-Veltman-Zakharov (vDVZ) discontinuity \cite{vanDam:,Zakharov}. Boulware and Deser also showed that a generic extension of the Fierz-Pauli theory to curved backgrounds will contain ghost instabilities \cite{Boulware,Boulware2}.  A theory  with no ghost in the decoupling limit, or at least up to quartic order away from the decoupling limit, has been constructed de Rham et al. constructed in \cite{deRham3,deRham}. Later, 
a generalisation to a ghost-free non-linear massive gravity action for all orders was proposed in \cite{HassanF}. \\ 

 On the experimental ground, recent observations by LIGO did not rule out the possibility of non zero mass, and put an upper bound on graviton mass: $m_{g}<1.2\times 10^{-22} eV/c^{2}$ \cite{Abbott,Abbott2}. For more details about the theory of massive gravity, we refer the reader to \cite{deRham2}.  \\ 
 

Black holes thermodynamics  have a long history, starting with Hawking, Carter, Bardeen and Page \cite{Bardeen:1973gs,Hawking:1982dh} as early as 1973, and is by now a known rich area. In particular, the extended phase space version of black holes thermodynamics has attracted attention during the last decades \cite{Kastor:2009wy,Dolan:2011xt}. Extension of this fascinating subject to the other theories of gravity via different approaches is relatively rather new \cite{Kubiznak,Belhaj,Chen,Mann}, especially in the massive gravity context \cite{Cai2014,Xu}. The key idea  originates from the identification of the cosmological constant $\Lambda$ as a thermodynamical pressure. This entails a phase structure like van der Waals one. In this sense, a variety of technics have born to consolidate this analogy, such as  thermodynamical geometry \cite{Belhaj:2015uwa,Chabab:2015ytz},  quasi-normal modes \cite{Chabab1,Chabab2} and non-local observables \cite{ElMoumni:2018fml,Belhaj:2019idh}. More recently, many endeavors have been performed to achieve connection between the thermodynamic behaviors and the geodesics of test particles \cite{Wei,Wei2,Bhamidipati}.  The purpose of the present paper is to study the unstable circular orbits of photons in the vicinity of charged black holes in asymptotically AdS space in the theory of massive gravity.

This paper is organized as follows. In the next section, we present a concise review of the critical behavior associated with  massive gravity charged black holes in AdS space. In section \ref{sec3},  we first establish the equation of motion for a free photon, propagating in the equatorial plane around RN-AdS black hole in massive gravity. Then, the conditions of the unstable circular photon orbits are obtained from the effective potential. In section \ref{sec4}, we discuss how the behaviors of radius and  minimal impact parameter of unstable circular photon orbits can reveal the thermodynamic phase stricture. In section \ref{sec5}, we investigate the unstable circular orbits behavior at the second order phase transition. The critical exponent associated with radius and minimal impact parameter are also calculated. The last section is devoted to our conclusion.

\section{Thermodynamics of charged AdS black holes in massive gravity}\label{sec2}

  In a 4-dimensional massive gravity model with negative cosmological constant,  a charged black hole solution has been obtained in \cite{Vegh}. The latter arising in the generalization of the theory introduced in \cite{HassanF2} is shown to be ghost free for an arbitrary non-dynamical reference metric \cite{HassanF3},  and read as  \cite{Zeng,Hendimassiv, HendiEBI},
\begin{equation}\label{lineelement}
ds^{2}=-f(r)dt^{2}+f(r)^{-1}dr^{2}+r^{2}d\theta^{2}+r^{2}\sin \theta d\phi^{2},
\end{equation}
with
\begin{equation}\label{metric}
f(r)=1-\frac{2 M}{r}+\frac{Q^{2}}{4 r^{2}}+\frac{r^{2}}{l^{2}}+a r+b,
\end{equation}
 $c_{0}$, $c_{1}$ and $c_{2}$ are constants and $m_{g}$  the graviton mass. $l$ represents the AdS length scale set by the presence of the cosmological constant. For commodity, we introduce in the solution the following denotations 
\begin{equation}\label{denotation}
a=m_{g}^{2} \frac{c_{0} c_{1}}{2} \ ,\  b=m_{g}^{2} c_{0}^{2}c_{2}.
\end{equation}

Following \cite{Kastor}, we may regard the cosmological constant as a variable and treat it as a dynamical pressure of black hole,
\begin{equation}\label{pressurecosmolog}
P=-\frac{\Lambda}{8\pi}=\frac{3}{8\pi l^{2}}.
\end{equation}
At the horizon $r_{h}$, from the condition $f\left(r_{h}\right)=0$, we obtain the black hole mass as, 
\begin{equation}\label{m}
M=\frac{12 a S^{3/2}+4 \sqrt{\pi } S (3 b+8 P S+3)+3 \pi ^{3/2} Q^2}{24
	\pi  \sqrt{S}},
\end{equation}
where, $S=\pi r_{h}^{2}$, is the entropy of the black hole. Also, since the Hawking temperature of the black hole is related to the surface gravity $T=\frac{\kappa}{2 \pi}$ with $\kappa=\frac{f'(r_{h})}{2}$, we can derive the thermodynamical equation of state,
\begin{equation}\label{Tstateequation}
T=\frac{8 a S^{3/2}+4 \sqrt{\pi } S (b+8 P S+1)-\pi ^{3/2} Q^2}{16 \pi  S^{3/2}}.
\end{equation}
The other conjugate quantities of the intensive parameters $P$, $Q$, $a$ and $b$ are given by,
\begin{equation}\label{conjugatequantitiesV}
V=\frac{4 S^{3/2}}{3 \sqrt{\pi }}, \ \Phi=\frac{\sqrt{\pi } Q}{4 \sqrt{S}},\  \mathcal{A}=\frac{S}{2 \pi },\  \mathcal{B}=\sqrt{\frac{S}{4 \pi }},
\end{equation}
respectively.  All the above quantities satisfy the first law of the black hole thermodynamics, 

\begin{equation}\label{firstlaw}
dM=TdS+\Phi dQ+VdP+\mathcal{A} da+\mathcal{B} db.
\end{equation}

Notice that the latter equation has been extended to include the massive gravity terms in our context,

Using  the scaling argument \cite{Kastor}, one obtain the Smarr formula  as:
\begin{equation}\label{Smarr}
M=2T S-2V P+\Phi Q-\mathcal{A} a.
\end{equation}
It is worth to note here that  $b$ does not appear in the Smarr formula, since its scaled weight vanishes \cite{Hendi}. One can also see that our $b$ in the metric function, Eq. \eqref{metric}, is a constant term with no thermodynamical contribution. 

Next we focus on the study of criticality. To this end, we first have to find out the
inflection point of $T(r_{h})$, by solving the following equations: 
\begin{equation}\label{critical point equation}
\left.\frac{\partial T}{\partial S}\right|_{S_{c},P_{c}}=\left.\frac{\partial^2 T}{\partial S^2}\right|_{S_{c},P_{c}}=0,
\end{equation}
 The critical point occurs when,
\begin{equation}\label{critical point pc}
P_{c}=\frac{(b+1)^2}{24 \pi  Q^2},\ S_{c}=\frac{3 \pi  Q^2}{2 (b+1)},\ T_{c}=\frac{a}{2 \pi }+\frac{2 (b+1)^{3/2}}{3\sqrt{6} \pi  Q}.
\end{equation}

Another key quantity characterizing the critical phenomena is the Gibbs free energy $G=M-TS$, given by
\begin{equation}\label{Gibbs}
G=\frac{4 S (3 b-8 P S+3)+9 \pi  Q^2}{48 \sqrt{\pi } \sqrt{S}}.
\end{equation}

\begin{figure}[h]
	\begin{center}
		\begin{tabbing}
			\hspace{8.5cm}\=\kill
		\includegraphics[width=8cm,height=4.cm]{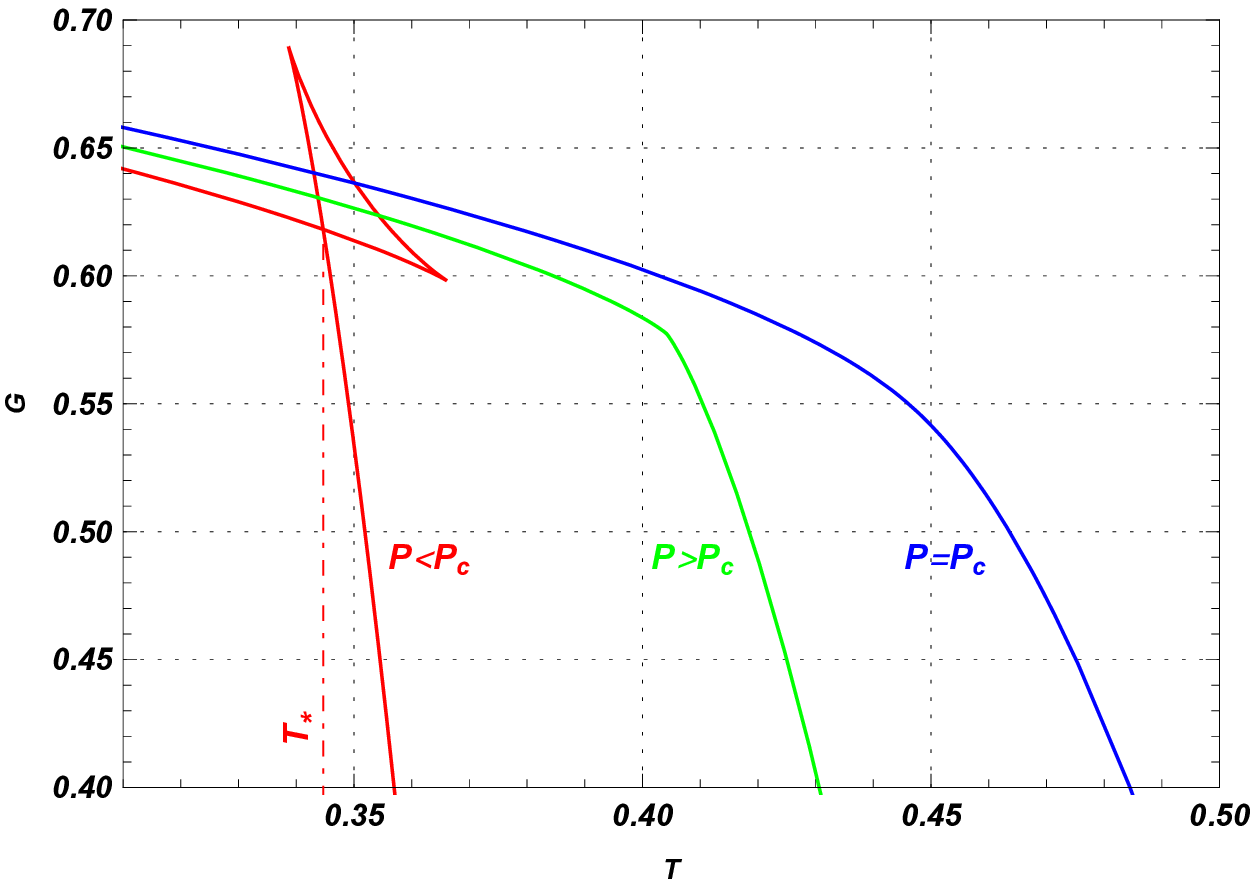}\>	\includegraphics[width=8cm,height=4.cm]{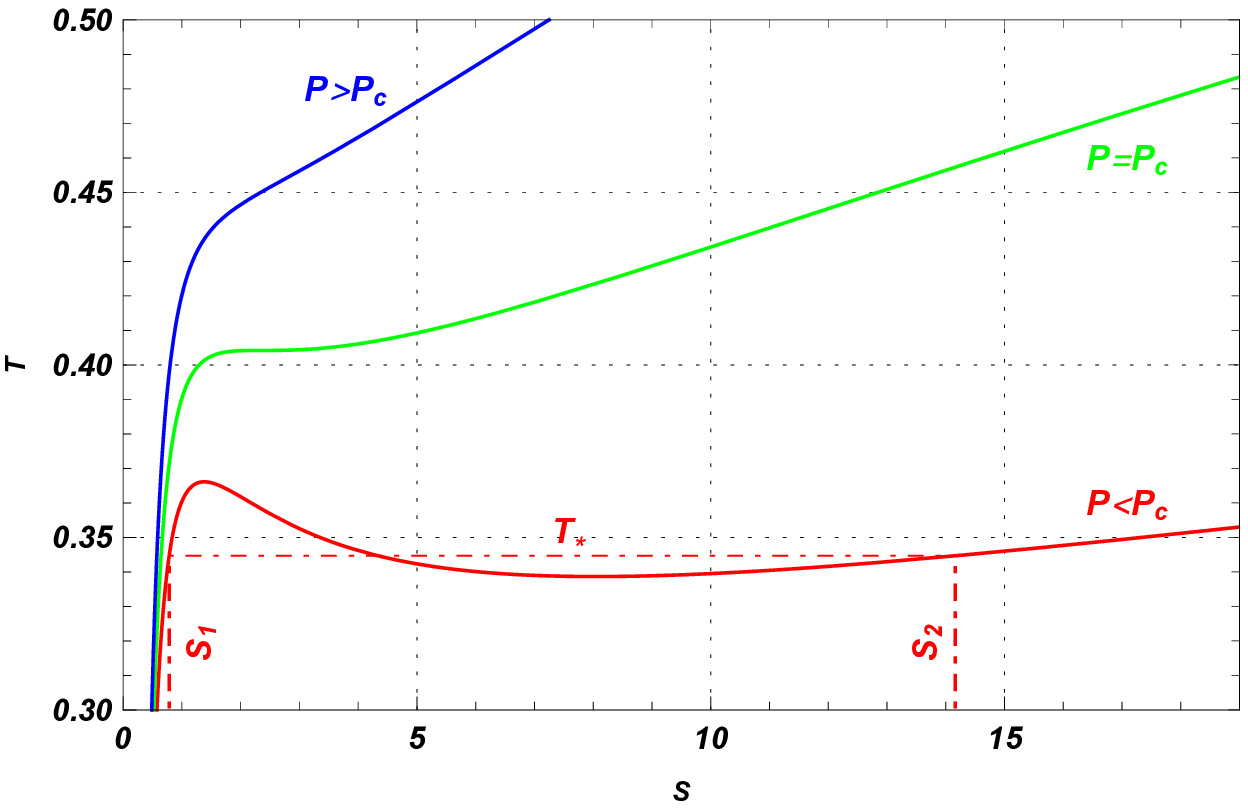}
		\end{tabbing}
	\end{center}
	\caption{Left: the Gibbs free energy as function of the temperature. Right: the Hawking temperature as function of the black holes entropy for various pressures. We set $a=1,\ b=1\ \text{and}\ Q=1$.} 
	\label{figGT}
\end{figure}

Now, having calculated the essential of thermodynamic quantities, we turn our attention to analyzing the corresponding phase transition. For this, we plot in figure \ref{figGT}, the variation of the temperature and the Gibbs free energy as a function of the entropy and temperature, respectively.

It can be seen from the left panel that the $T-S$ diagram is exactly the same as the $T-S$ criticality of the van der Waals liquid-gas system. As discussed in \cite{Mann,Chabab:2015ytz},the $G-T$
 diagram develops a swallow tail behavior when $P<P_c$
which indicates a first order phase transition as expected. This feature disappears above the critical pressure and thus no phase transition small/large black holes shows up. However, the phase transition  becomes instead a second order at $P=P_{c}$. To describe the  phase transition one has to replace the oscillating part between $S_{1}$ and $S_{2}$ of the $T-S$ diagram by an isotherm line, $T=T_{*}$, according to Maxwell equal area law:
\begin{equation}\label{areal law}
	\int_{S_{1}}^{S_{2}} TdS=T_{*}\left(S_{2}-S_{1}\right).
\end{equation}
This prescription reflects the fact that both phases have the same Gibbs free energy at the phase transition. Next, we will employ all these informations to revisit the phase transitions by means of the photon orbits formalism.

\FloatBarrier
\section{Geodesic equations of motion}\label{sec3}
In this section, we consider the motion of a free photon propagating in the background geometry of the metric given by Eq. \eqref{metric}.  Its dynamics  can be treated via the Lagrangian formalism as, 
\begin{equation}\label{largangiian}
\mathcal{L}=\frac{1}{2}g_{\mu\nu}\frac{d x^{\mu} }{d \sigma}\frac{d x^{\nu} }{d \sigma},
\end{equation}
with $x^{\mu}=\left(t,r,\theta,\phi\right)$ and  $\sigma$ is an affine parameter. The components of 4-momentum  are  related to the coordinate $x^\mu$ by
\begin{equation}\label{momentum}
p_{\mu}=\frac{\partial \mathcal{L}}{\partial \dot{x^{\mu}}},
\end{equation}
Here, the dot denotes differentiation with respect to the affine parameter $\sigma$. Without loss of generality, we will restrict our analysis to the equatorial geodesics by setting  $\theta=\frac{\pi}{2}$ and  $\dot{\theta}=0$.  The spacetime symmetries (i.e the metric) has two Killing vectors which imply the existence of two constants of motion
\begin{equation}
p_{t}=-E\  \text{and} \ p_{\phi}=L.
\end{equation}

where $E$ and $L$ represent the energy and angular momentum, respectively \cite{shapiro, Chandrasekhar}.
Using the Euler-Lagrange equations with respect to $t$ and $\phi$, one obtain the equations,
\begin{equation}\label{motioneuqationpt}
\frac{d t}{d \sigma}=\frac{E}{f(r)},
\end{equation}
\begin{equation}\label{motioneuqationpphi}
\frac{d \phi}{d \sigma}=\frac{L}{r^{2}}.
\end{equation}
Next, we introduce the effective potential by recalling the normalisation condition for the null geodesics $g_{\mu\nu}\dot{x^{\mu}}\dot{x^{\nu}}=0$. Indeed, under this constraint,  the radial motion takes the following form,
\begin{equation}\label{Hamiltonien}
\dot{r}^{2}+V_{eff}(r)=0,
\end{equation}
where the effective potential $V_{eff}$  reads as:
\begin{equation}\label{veff}
\frac{V_{eff}(r)}{E^{2}}=\xi^{2}\frac{f(r)}{r^{2}}-1.
\end{equation}
where $\xi=\frac{L}{E}$ denotes the impact parameter. The photons motion is confined to the region where $V_{eff}\le 0$ \cite{Wei}.

 Because the most important class of orbits are circular orbits, we will focus on the study of the unstable circular  orbits  in which a photon coming from infinity approaches the turning points with zero radial velocity. The radius $r_{0}$ of the unstable circular orbit can be found by simultaneously solving  the equations
\begin{equation}\label{unstablecircular}
  V_{eff}(r_{0})=0,\ \ \left.\frac{\partial V_{eff}(r)}{\partial r}\right|_{r_{0}}=0,\ \ \left.\frac{\partial^{2} V_{eff}(r_{0})}{\partial r^{2}}\right|_{r_{0}}<0.
\end{equation}
Also, One can simply  use Eq. \eqref{veff} to obtain the expression of the minimal impact parameter corresponding to the unstable circular orbits:
\begin{equation}\label{criticalimpact}
\xi_{0}=\frac{r_{0}}{\sqrt{f(r_{0})}}.
\end{equation}

\begin{figure}[h]
	\begin{center}
		\includegraphics[width=9cm,height=6cm]{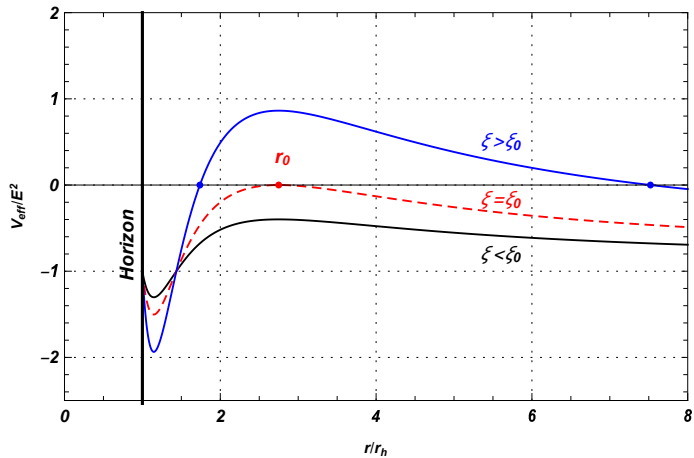}
	\end{center}
	\caption{The effective potential profile for a free photon orbiting around a charged AdS black hole in massive gravity with $S=0.2$, $P=0.05$, $a=1,\ b=1\ \text{and}\ Q=1$. The red and blue dots represent the turning points.} 
	\label{figveff}
\end{figure}
The effective potential is depicted schematically  in figure \ref{figveff}. Several important aspects of the motion can be deduced from this figure. In fact, the nature of the orbit will  be governed by the turning points determined by $V_{eff}(r) = 0$. In particular,  if the photon fall from infinity with $\xi>\xi_{0}$, it will reach the turning point  and then scattered back out to larger $r$. This is similar to the hyperbolic orbits in Newtonian gravity. However, the photon will be captured by the black hole if $\xi<\xi_{0}$, where there is no turning point, so that the capturing cross section for photons from infinity is given by,
\begin{equation}\label{crosssetion}
\sigma_{capture}=\pi \xi_{0}^{2}=\frac{\pi r_{0}^{2}}{f(r_{0})}.
\end{equation}
 Finally, the case $\xi=\xi_{0}$  is the critical impact parameter separating capture from scattering orbits region. After showing the main photons dynamical concepts in such black hole configurations, we will probe in the next section, how to establish a direct link between this dynamics and thermodynamics.


\section{Thermodynamic phase transition  from unstable circular photon orbits}\label{sec4}

Our main objective here is to elaborate a connection between the unstable circular photon orbits with the thermodynamic phase transition of charged AdS black holes in massive gravity.  The expected outcome is, on one hand, a generalization of the study of \cite{Wei} and, on the other hand, to pave the way to a  novel understanding of the thermodynamic behavior of RN-AdS black holes in massive gravity. To this end, we first substitute 
 Eq. \eqref{metric} in Eq. \eqref{veff} under the requirement $V'_{eff}(r_{0})=0$ for unstable circular orbits, then we get the following constraint: 
 \begin{equation}\label{unstablecircular2}
 a r_0^3+2( b+1) r_0^2-6 M r_0+Q^2=0.
 \end{equation}
As in the RN-AdS black holes, the instability conditions given by Eq. \eqref{unstablecircular} are not explicitly affected by the AdS radius. Hence, for a fixed black hole mass, $V'_{eff}(r)$ remains the same  either for asymptotically de-Sitter space or for flat space.  To obtain the unstable circular radius $r_{0}$, we need to combine Eq. \eqref{m} with Eq. \eqref{unstablecircular2} and solve a third degree equation.  Then we inject the solution into Eq. \eqref{criticalimpact} to extract the impact parameter $\xi_{0}$. The resulting solution can be expressed in terms of the parameters $P$, $S$, $Q$, $a$  and $b$.  Because of its lengthy expression, we will not show it here.

\begin{figure}[h]
	\begin{center}
\begin{tabbing}
	\hspace{8.5cm}\=\kill
		\includegraphics[width=8cm,height=4.5cm]{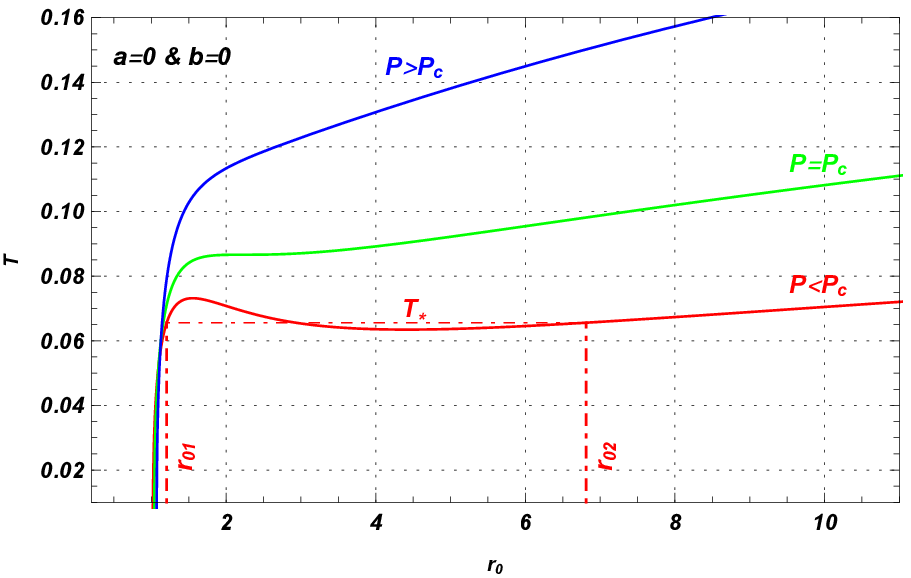}	\> \includegraphics[width=8.15cm,height=4.5cm]{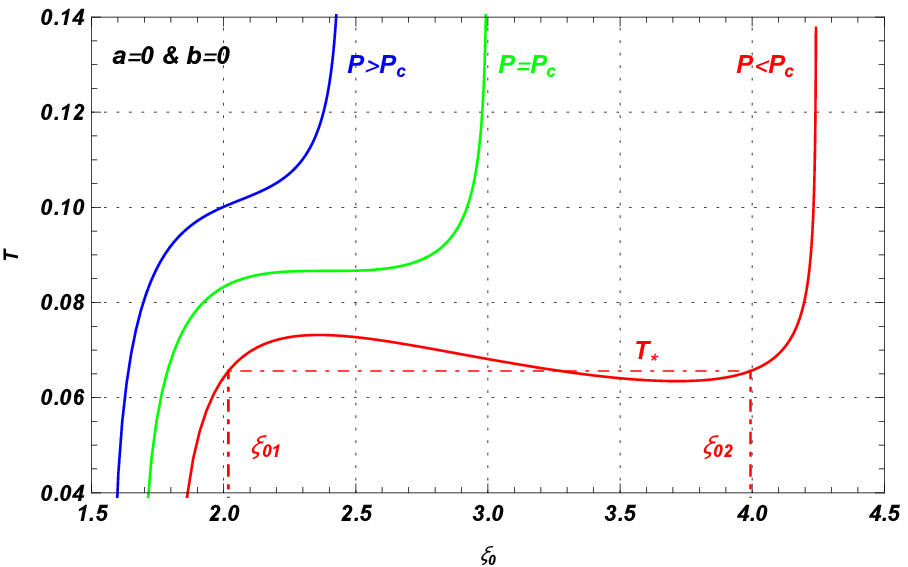}\\ 
\includegraphics[width=8cm,height=4.5cm]{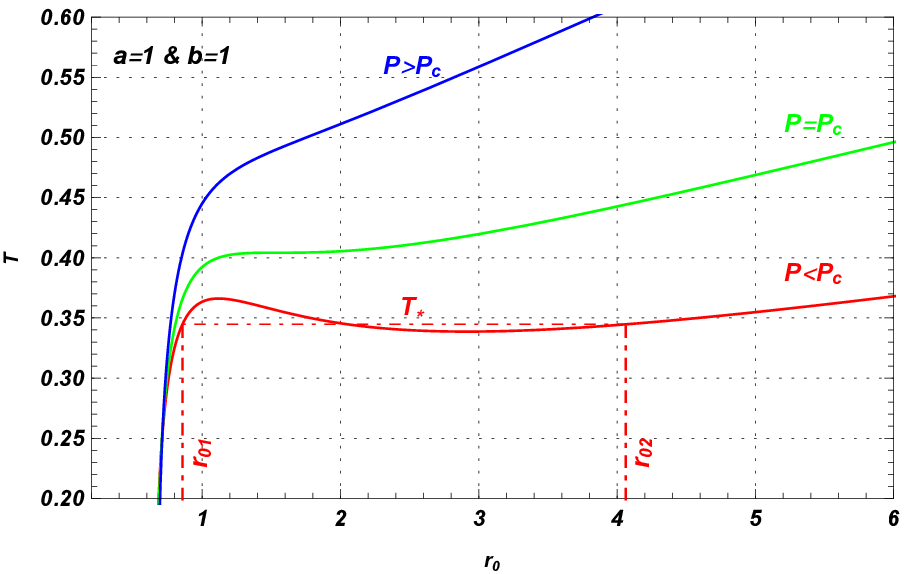}	\> \includegraphics[width=8cm,height=4.5cm]{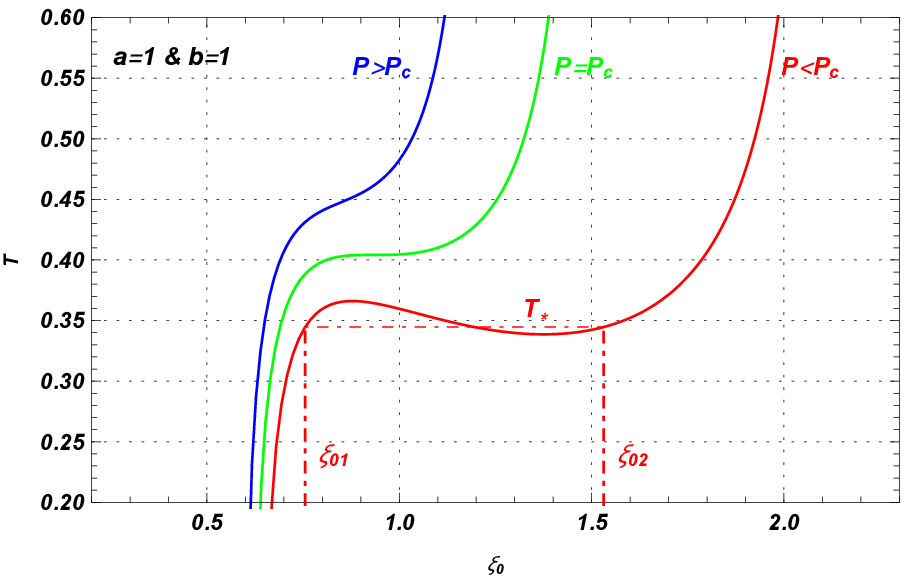} \\ 
	\includegraphics[width=8cm,height=4.5cm]{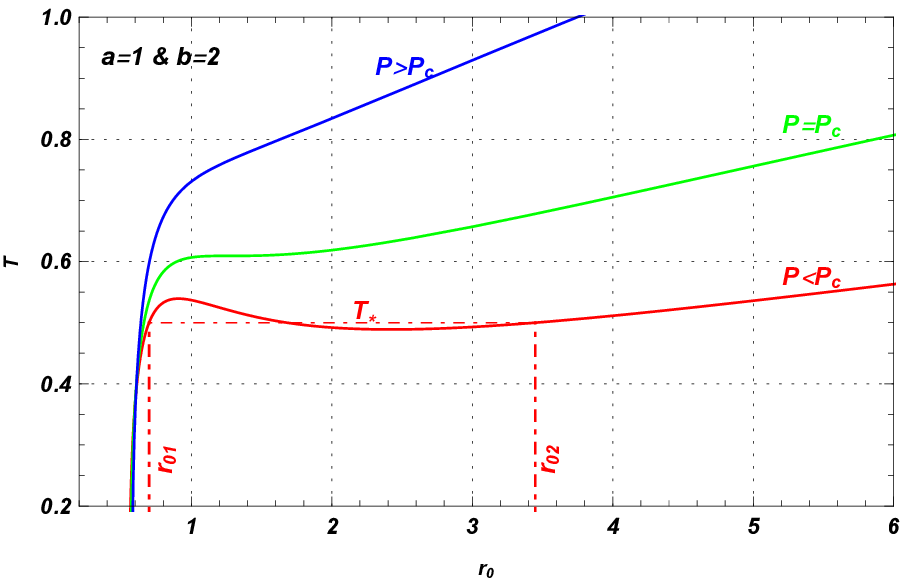}	\>	\includegraphics[width=8cm,height=4.5cm]{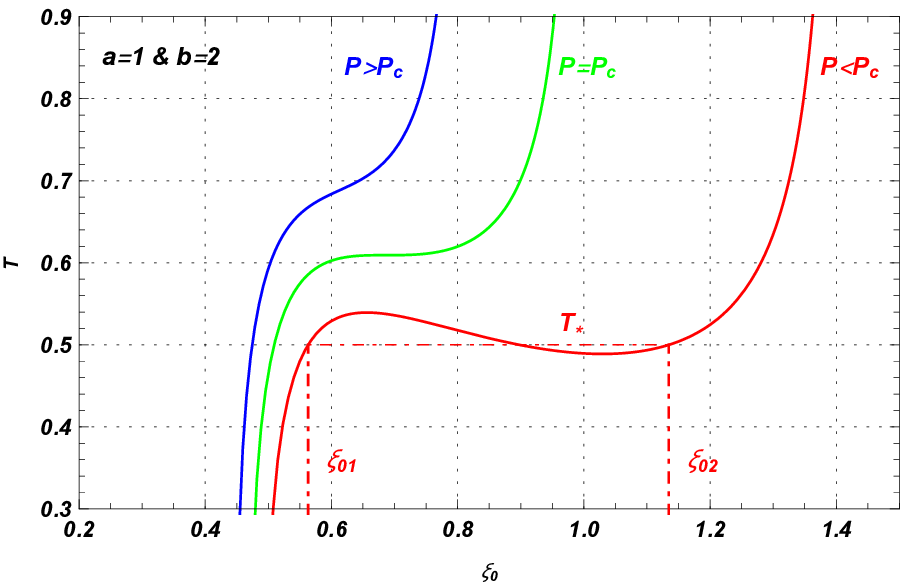} \\ 
	\includegraphics[width=8cm,height=4.5cm]{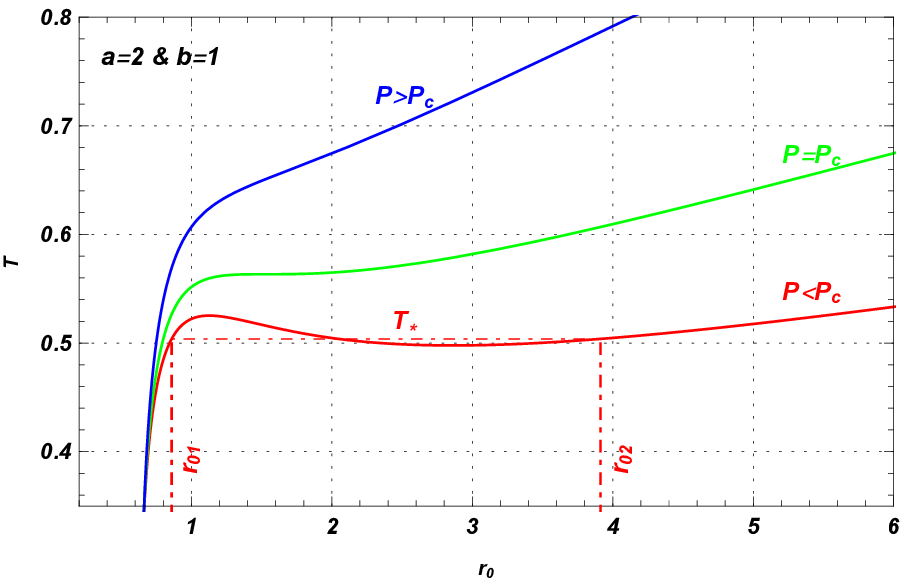}	\> \includegraphics[width=8.15cm,height=4.5cm]{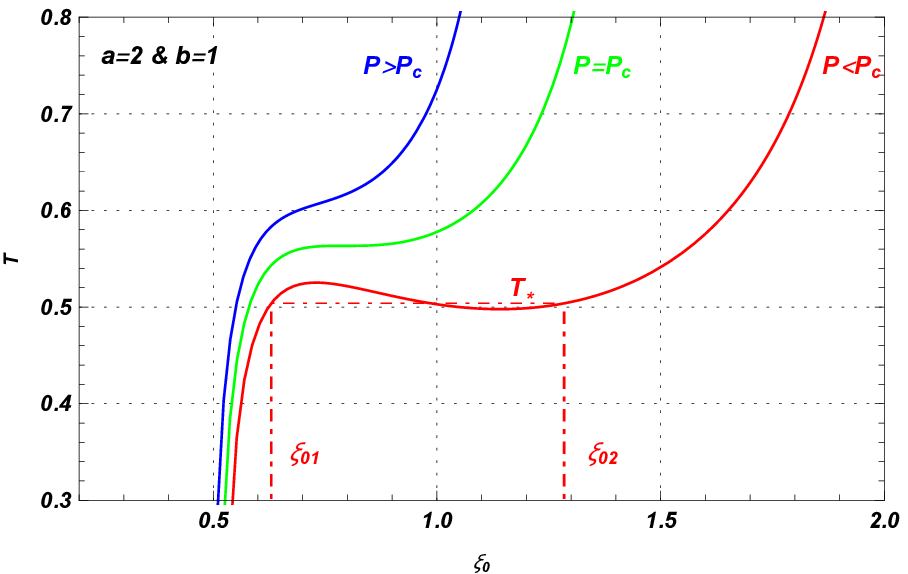}
\end{tabbing} 
	\end{center}
\caption{$T-r_{0}$ diagram (left) and $T-\xi_{0}$ diagram (right) in the isobaric process, for different values of massive gravity parameters $a$ and $b$, where we set $Q=1$.} 
\label{figr0}
\end{figure}

In each panel of Fig. \ref{figr0}, we consider  the isobaric process for which we plot the temperature as a function of  the  unstable circular radius and minimal impact parameter for different values of massive gravity terms ($a$ and $b$). We  see that  the $T-r_{0}$ and $T-\xi_{0}$ behaviors do not change with $a$ and $b$ variations.  For the special  case of RN-AdS black holes, where $a=0$ and $b=0$, we reproduce the features of the unstable circular orbits derived in \cite{Wei} in four dimension spacetime.

Moreover, by analyzing  Fig. \ref{figr0}, one also observe that the all panels look like the right panel of figure \ref{figGT} illustrating $T-S$ behavior. Indeed, for $P<P_{c}$, whatever the chosen values of massive gravity parameters ($a$ and $b$), we observe an oscillating curve which indicates that there is a first-order phase transition between two extremal points corresponding to the small black hole $r_{01}$  ($\xi_{01}$) and large black hole $r_{02}$ ($\xi_{02}$) respectively. However, for $P=P_{c}$,  an inflexion point shows up to signal  the second order phase transition occurrence. Besides, above the critical point, the  behavior of the unstable circular orbits do not show oscillating line, and the temperature is a monotonically increasing function of $r_{0}$  (or $\xi_{0}$): no phase transition can happen. We keep in mind that one can find the coexistence temperature $T_{*}$  and the extremal points from the $G(T)$ and $T(S)$ equations respectively.

In summary, the behaviors of $r_{0}$ and $\xi_{0}$ confirm the same phase transition picture as that from the entropic side. Thereby, the photon in orbits is definitely an efficient tool to  probe easily the phase transition between small and large black holes in massive gravity.

\FloatBarrier
\section{Null geodesics and critical exponents}\label{sec5}

In this section, let us examine the behavior of unstable circular orbits of photons  near the second order phase transition. Thus, we need to  calculate  the critical exponent associated with the radius and the minimal impact parameter. For this, we introduce  
the reduced  temperature, the radius $r_{0}$ and the minimal impact parameter $\xi_{0}$ as follow:
 \begin{equation}\label{reduced T}
\tilde{T}=\frac{T}{T_{c}}=\frac{3 \sqrt{\frac{3}{2}} Q \left(8 a S^{3/2}+4 \sqrt{\pi } S (b+8 P S+1)-\pi ^{3/2} Q^2\right)}{4 S^{3/2} \left(3 \sqrt{6} a Q+4 (b+1)^{3/2}\right)},
\end{equation}
 \begin{equation}\label{reduced r0xi0}
\tilde{r}_{0}=\frac{r_{0}}{r_{0c}},\ \tilde{\xi}_{0}=\frac{\xi_{0}}{\xi_{0c}},
\end{equation}
where $r_{0c}$ and $\xi_{0c}$ are the values of unstable circular radius and the minimal impact parameter at the critical point, respectively. They are obtained by replacing $S$ by $S_{c}$  and $P$ by $P_{c}$ in their corresponding expressions. Then, we determine the width of the coexistence lines in both  $\tilde{T}-\tilde{r}_{0}$ and $\tilde{T}-\tilde{\xi}_{0}$ diagrams as, 
 \begin{equation}\label{delta r0xi0}
\Delta \tilde{r}_{0}=\tilde{r}_{02}-\tilde{r}_{01},\ \Delta \tilde{\xi}_{0}= \tilde{\xi}_{02}- \tilde{\xi}_{01}.
\end{equation}

\begin{figure}[h]
	\begin{center}
		\includegraphics[width=8cm,height=5cm]{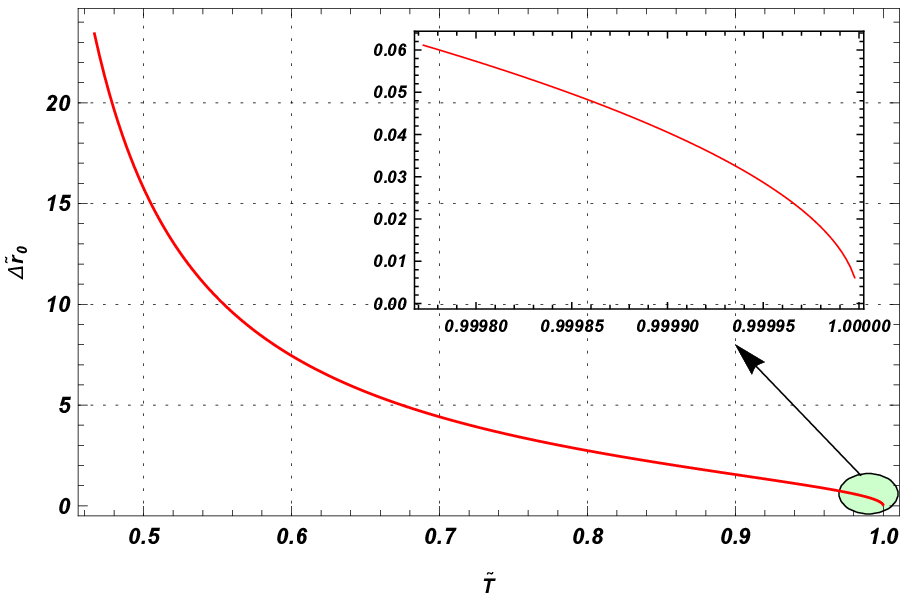}
		\includegraphics[width=8cm,height=5cm]{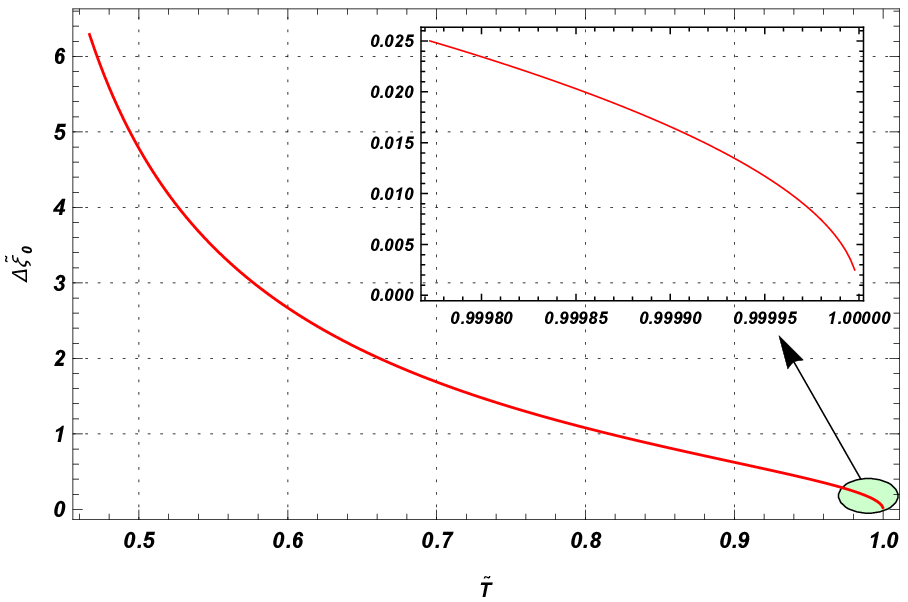} 
	\end{center}
	\caption{$\Delta \tilde{r}_{0}$ (left panel), $\Delta \tilde{\xi}_{0}$ (right panel) as a function of coexistence temperature. We set $a=1,\ b=1\ \text{and}\ Q=1$.} 
	\label{figdelta0}
\end{figure}

In the left panel of Fig. \ref{figdelta0}, we plot the difference $\Delta \tilde{r}_{0}$ as a function of   $\tilde{T}$, which  allows us to identify  $\Delta \tilde{r}_{0}$ as an order parameter for the small-large black holes  phase transition. Indeed, as expected, the width of the coexistence line decreases with the reduced temperature, has  non zero values at the first order phase transition,  and tends to zero at the critical point. Furthermore, by the zooming near the second order  phase transition in Fig. \ref{figdelta0},  one can see that the concavity of the graph is reversed, namely:  It is convex at the first order phase transition and becomes concave when temperature approaches $T_{c}$. The behavior of $\Delta \tilde{\xi}_{0}$ in the right panel  of Fig.  \ref{figdelta0}  illustrates similar features to those of $\Delta \tilde{r}_{0}$.

\begin{figure}[h]
	\begin{center}
		\includegraphics[width=8cm,height=5cm]{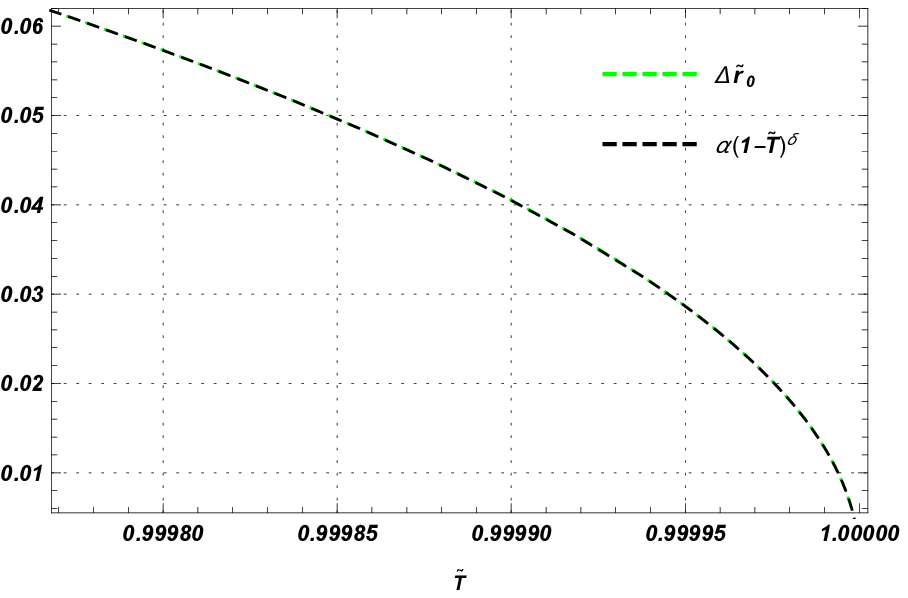}
		\includegraphics[width=8cm,height=5cm]{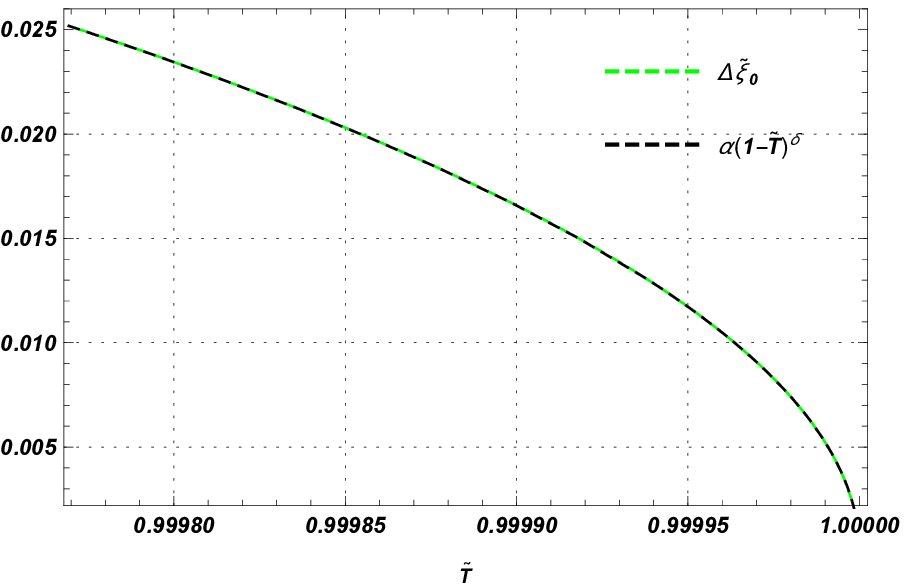} 
	\end{center}
	\caption{Fitting curves, $\Delta \tilde{r}_{0}$ (left) and $\Delta \tilde{\xi}_{0}$ (right) versus $\tilde{T_{*}}$ near the critical temperature. We set $a=1,\ b=1\ \text{and}\ Q=1$.} 
	\label{figfitdelta}
\end{figure}

Therefore, it will be interesting to check the critical exponent describing the universal behavior of the reduced differences $\Delta \tilde{r}_{0}$ and $\Delta \tilde{\xi}_{0}$ around the second order phase transition. To this end, we need to fit our numerical data. Thus, by looking for the following form,
 \begin{equation}\label{fit}
	\sim \alpha \left(1-\tilde{T}\right)^{\delta},
\end{equation}
we can  identify the power law functional relationship between the two reduced quantities of the unstable circular orbits and the reduced temperature near $T_{c}$. Table \ref{table fit} displays the results obtained for $\alpha$ and $\delta$ parameters for different values of massive gravity terms ($a$ and $b$). In Fig.  \ref{figfitdelta}, we display a comparison between the fitting curves and $\Delta \tilde{r}_{0}$ (and $\Delta \tilde{\xi}_{0}$) presented in the zoom panels of Fig.  \ref{figdelta0}.

{\rowcolors{3}{gray!15}{}
\begin{table}[h]
\begin{center}
	\begin{tabular}{cccccc} 
		\hline\hline
\rowcolor{white!0} \multirow{2}{*}{}  &  \multirow{2}{*}{} & \multicolumn{2}{c}{$\Delta \tilde{r}_{0}$}   &  \multicolumn{2}{c}{$\Delta \tilde{\xi}_{0}$}    \\ 
 \cline{3-6}    \rowcolor{white!0}
 $a$ & $b$ & $ \alpha$ & $\delta$  & $ \alpha$ & $\delta$ \\ 
\hline
0	& 0  & 3.47293  & 0.500258& 1.20058& 0.500213\\  
1	& 1 &  4.05847 &  0.500192 & 1.66049 & 0.500174\\  
1	& 2 & 3.78479 & 0.500206  &  1.46958 & 0.500191\\  
1	& 3 &  3.67083 & 0.500216 & 1.38155 & 0.500198\\  
1	&4  & 3.61196 & 0.500224 & 1.33271 & 0.500203\\  
2	&1  & 4.62966  & 0.500179 & 2.01595 & 0.500148\\  
3& 1 & 5.15089 & 0.500186  & 2.31723 & 0.500144\\  
4	& 1 &  5.62841 &  0.500188 & 2.58336  & 0.500138\\ \hline \hline
\end{tabular} 
\caption{The critical exponent $\delta$ and the proportionality factor  $\alpha$ for the unstable circular photon orbits  for different values of $a$ and $b$,  with $Q=1$.}\label{table fit}
\end{center}
\end{table}
}

Our results confirm clearly, within the used numerical accuracy, that the unstable circular photon orbits (with $\Delta  \tilde{r}_{0}$ and $\Delta \tilde{\xi}_{0}$) around RN-AdS black hole in massive gravity present a universal critical exponents  equal to $1/2$.  Furthermore, one can also use the derived formula given by Eq. \eqref{fit} to check the concavity of the graphs shown in Fig. \ref{figdelta0} in the vicinity of the critical temperature. Indeed, since
\begin{equation}\label{concavity}
\frac{d^{2}\Delta\tilde{r}_{0}}{d \tilde{T}^{2}}=-\alpha \delta \left(1-\delta\right)\left(1-\tilde{T}\right)^{\delta-2}<0,
\end{equation}
those graphs must be strictly concave when $T\rightarrow T_{c}$. Hence, the detection of concavity change can help to uncover the thermodynamic criticality.

At last, these results confirm that the studies of unstable circular orbits of photons can be a useful tool to reveal the Van der Waals-like phase transition between small and large black holes in AdS massive gravity background. 

\FloatBarrier
\section{Conclusion}
In this work we have explored the unstable circular orbits of photons around charged AdS black holes in massive gravity, as a new bridge to the Van der Waals-like phase structure. In particular, we have extended the work of \cite{Wei} by considering the massive gravity background.  Foremost, we have reviewed briefly the thermodynamic  behavior of the selected class of black holes. Then using the Lagrangian formalism   we  have calculated the equation of motion that govern a photon motion in the equatorial plane. We have also presented two key quantities to this study, namely the radius $r_{0}$  and minimal impact parameter $\xi_{0}$ corresponding to the unstable circular orbits. The plots of these two quantities as a function of temperature have shown an oscillatory curves, revealing the presence of a  coexistence region for small/large black hole, and indicating that the thermodynamic phase transitions has taken place. On the other hand, at the second order phase transition, we have examined the order parameters for small-large black hole phase transition in $\Delta \tilde{\xi}_{0}-\tilde{T}$ and $\Delta \tilde{r}_{0}-\tilde{T}$ diagrams,  and as a result, we found that the concavities of both plots are reversed  near the critical point. 

Furthermore, by means of a numerical fit to a power law relationship between fitting $\Delta \tilde{r}_{0}$ and $\Delta \tilde{\xi}_{0}$, we have identified a universal critical component $\delta=1/2$. Hence, this new approach leads to a set of critical exponents similar to those of van der Waals fluid system,   providing a new insight into the phase picture from the point view of photon trajectories.

At the end, we plan to extend the present study further by considering other gravity configurations or others kinds of test particule, especially the neutrinos.

\end{document}